\documentstyle[aps,prb,eqsecnum,epsf]{revtex}

\newcommand{\br}{{\bf r}}

\newcommand{\be}{\begin{equation}}
\newcommand{\ee}{\end{equation}}

\begin{document}

\title{Recent Progress in the Computational Many-Body Theory of Metal
Surfaces}
\author{Maziar Nekovee$^1$ and J. M. Pitarke$^{2,3}$} 
\address{$^1$Centre for Computational Science, Queen Mary and
Westfield College, University of London,\\ 
Mile End Road, London E1 4NS, UK\\
$^2$Materia Kondentsatuaren Fisika Saila, Zientzi
Fakultatea, Euskal Herriko  Unibertsitatea,\\ 644 Posta kutxatila, 48080
Bilbo,
Basque Country, Spain \\
$^3$Donostia International Physics Center (DIPC) and Centro Mixto
CSIC-UPV/EHU,\\
Donostia, Basque Country, Spain}

\date{\today} 

\maketitle

\begin{abstract}
{In this article we describe recent progress in the computational 
many-body theory of metal surfaces, and focus on current techniques
beyond
the local-density approximation of density-functional theory. We
overview
various applications to ground and excited states. We discuss the
exchange-correlation hole, the surface energy, and the work
function of 
jellium surfaces, as obtained within the random-phase approximation, a
time-dependent density-functional approach, and quantum Monte Carlo
methods. We
also present a survey of recent quasiparticle calculations of unoccupied
states
at both jellium and real surfaces.}
\end{abstract}


%\tableofcontents

\section{Introduction}
\label{sec:intro}

Prerequisite for calculating the electronic properties of solid surfaces 
is a practical procedure for the treatment of electron-electron (e-e) 
interactions in large, strongly inhomogeneous electron systems. Over the
last
three decades, almost all first-principles electronic-structure
calculations have
been carried out within density-functional theory (DFT)\cite{hk}, which
solves the many-electron problem by introducing an auxiliary
non-interacting
system with the same electron density\cite{Sham}. Although DFT is a
formally
exact theory, it relies on the knowledge of the so-called
exchange-correlation
(xc) energy as a functional of the electron density. The best known
approximation for this functional is the local-density approximation
(LDA)\cite{Sham}, which has proved remarkably successful in various
applications to
ground-state properties of solid surfaces\cite{Inglesfield}. 
However, this simple
approximation presents two important shortcomings. First, the LDA is
unable to
provide the highly accurate total energies that are required in the
application
of DFT to chemical reactions at surfaces. Secondly, it yields an
inaccurate
description of the long-range xc-hole density at solid surfaces, and
fails to
reproduce the correct image-like asymptotic behaviour of the surface
barrier,
which plays an important role in the interpretation of a number of
surface-sensitive experiments. Furthermore, the knowledge of the finite
quasiparticle lifetimes of unoccupied states is beyond the scope of DFT.

The aim of this article is to survey recent computational work on the
description of many-body effects in both ground- and excited-state
properties of metal surfaces. In section II we give a short overview
of the
state-of-the-art computational many-body theories of the inhomogeneous
electron
gas: DFT, in the LDA and its various gradient-corrected forms, quantum
Monte
Carlo (QMC) methods, and Green function theory in the so-called GW
approximation.
In section III we focus on a description of ground-state properties of
metal 
surfaces, and we critically examine the performance of current many-body
schemes in the computation of two of the most fundamental magnitudes,
namely
the surface energy and the work function. In section IV we examine
quasiparticle calculations of the surface barrier and the binding
energies and
lifetimes of image-potential induced states. These so-called image
states\cite{Echenique} form a Rydberg-like series, which converges
towards the
vacuum energy, and their existence is perhaps the most striking
manifestation of
long-range many-body  effects at surfaces. We conclude this article in
section V
with some perspectives for future research in this area.

We use Hartree atomic units throughout, unless otherwise specified. In
these
units, $e=m_e=\hbar=4\pi\epsilon_0=1$.

\section{Theory of the inhomogeneous electron gas}
\label{sec:inter}

Under the assumption that electronic and ionic degrees of freedom can be
decoupled, the problem of $N$ electrons in a solid can be described by
the following Hamiltonian:
\begin{equation}
\label{hamiltonian}
\hat H=\sum_{i=1}^N\left[-\frac{1}{2}\nabla^2_i+V_{ext}(\br_i)\right]+
\sum_{i<j}^N\frac{1}{|{\br}_i-{\br}_j|}. 
\end{equation}
Here ${\br}_i$ is the coordinate of electron $i$, and $V_{ext}({\br}_i)$
is the
external potential accounting for the interaction with the fixed atomic
nuclei.

\subsection{Density-functional theory}
\label{subsec:methods-dft}

DFT is a formally exact theory based on two theorems by Hohenberg and 
Kohn \cite{hk}.
The first theorem is extraordinary: it states that the ground-state
density
$n(\br)$ of a system of interacting electrons determines the
ground-state
properties of the system uniquely. The second theorem states that the
ground-state energy can be  obtained by minimising an energy functional
$E[\tilde n(\br)]$ of the electron density. The minimum value of
$E[\tilde n(\br)]$ is the
exact ground-state energy, and is attained when $\tilde n(\br)$ is the
exact
ground-state density $n(\br)$. 

In the Kohn-Sham formulation of DFT\cite{Sham}, the ground-state density
can be
obtained by calculating the eigenfunctions $\phi_i({\bf r})$ and
eigenvalues
$\varepsilon_i$ of non-interacting, single-particle equations, yielding
\begin{equation}
\label{eq:den}
n({\br})=\sum_{i=1}^N|\phi_i({\bf r})|^2,
\end{equation} 
and the energy functional is written in the form
\be\label{xc}
E[\tilde n(\br)]=T_0[\tilde n(\br)]+\int
d\br\,\tilde n(\br)\,V_{ext}(\br)+{1\over 2}\int
d\br\,{\tilde n(\br)\,\tilde n(\br')\over|\br-\br'|}+E_{xc}[\tilde
n(\br)].
\ee
Here $T_0[\tilde n(\br)]$ is the kinetic-energy functional for {\em
non-interacting} electrons with density $\tilde n(\br)$, and
$E_{xc}[\tilde n(\br)]$ is the so-called xc energy
functional. $E_{xc}$ contains all many-body effects beyond the Hartree
approximation, as well as the difference between the {\it actual}
kinetic-energy
functional of the interacting system and $T_0[\tilde n(\br)]$. By
minimising the
energy functional of Eq. (\ref{xc}), one concludes that the {\it exact}
density
is obtained by solving self-consistently the so-called Kohn-Sham
equations
\begin{equation}
\label{ks}
\left[-\frac{1}{2}\nabla^2+V_{ext}({\bf r})+\int d{\bf
r}'\,\frac{n({\bf
r}')}{|{\bf r}-{\bf r}'|}+V_{xc}({\bf r})\right]\phi_i({\bf r})=
\epsilon_i\phi_i({\bf r}),
\end{equation}
where $V_{xc}(\br)$ is the xc potential,
\begin{equation}
V_{xc}({\bf r})=\left[\frac{\delta E_{xc}[\tilde n({\bf
r})]}{\delta\tilde n({\bf
r})}\right]_{\tilde n(\br)=n(\br)}.
\end{equation}

The xc energy functional is not known and must be approximated. The
simplest
approximation for $E_{xc}[n(\br)]$ is the LDA,
\begin{equation}
E_{xc}^{\rm LDA}[n({\bf r})]=\int d{\bf r}\,n({\bf
r})\,\epsilon_{xc}^{unif}(n({\bf r})),
\end{equation}
where the actual xc energy per particle at point $\br$,
$e_{xc}(\br;[n(\tilde\br)])$, which is a functional of the density
$n(\tilde\br)$, has been replaced by the xc energy per particle of a
uniform
electron gas of density $n(\br)$,
$\epsilon_{xc}^{unif}(n(\br))$\footnote{Several  parameterisations of
$\epsilon_{xc}(n)$ exist, which are based on many-body QMC
simulations of the uniform electron gas\cite{ca}.}. 

For an electron gas of slowly varying density the
LDA may be corrected through
the introduction of gradient expansions, as suggested by Kohn and
Sham\cite{Sham}. Although the basic physics behind gradient expansions
is
correct, Langreth and Perdew\cite{lp} showed that it fails in most
practical
situations because of the spurious small-wave-vector contribution to the
Fourier
transform of the second-order density-gradient expansion for the xc hole
around
an electron. Via cutoff of this spurious contribution, generalised
gradient
approximations (GGA) were then
devised\cite{gga1,perdew_90,gga2,gga3,becke},
which approximate the xc energy per particle as a carefully chosen
nonlinear function of the electron density and its gradient. Recently, a
meta-generalised gradient approximation (meta-GGA) has been
developed\cite{meta},
which
makes use not only of the local density and its gradient but also of 
the orbital kinetic-energy density. For other kinds of improvements of
the LDA, including
the weighted density approximation (WDA)\cite{wd} and self-interaction
corrections (SIC)\cite{si1,si2}, we refer the reader to the review by
Perdew and
Kurth\cite{pk}.

\subsection{Quantum Monte Carlo methods}
\label{subsec:methods-qmc}

Unlike the density-functional approach, in which 
the ground-state density is the basic quantity, wave-function-based
quantum Monte
Carlo methods sample the full ground-state many-body
wave function of the system under consideration\cite{Ceperley1}.
There are many different QMC methods, but here we will be concerned
with only two of them: variational quantum Monte Carlo (VMC) and
fixed-node
diffusion quantum Monte Carlo (DMC), which have been shown to provide a
description of increasing sophistication of the properties of
zero-temperature
systems.

The VMC method is based on an ansatz for a parameterised trial many-body
wave
function. Expectation values are evaluated by the Metropolis Monte Carlo
techniques\cite{metro,qmc_review}, and the free parameters are then
varied in
order to optimise either the energy expectation value or the fluctuation
of the local energy \cite{qmc_review}.

We consider trial wave functions of the Slater-Jastrow type\cite{Jast}
\be\label{Jastrow}
\Psi=D\,\exp\left[\sum_{i=1}^N\chi_{\sigma_i}({\bf
r}_i)-\sum_{i<j}^Nu_{\sigma_i,
\sigma_j}(\br_i,\br_j)\right], 
\end{equation}
where $D$ is a Slater determinant of either Hartree-Fock (HF) or
LDA single-particle orbitals\footnote{$D$ is often split into electron
spin-up and spin-down Slater determinants, in order to reduce
computational
cost.}, $u_{\sigma_i,\sigma_j}$ is a two-body term correlating 
the motion of pairs of electrons, and $\sigma_{i}$ denotes the spin of
electron
$i$. 
The short-range form of $u_{\sigma_i,\sigma_j}$ is dictated by the well-known 
cusp condition
\cite{Kato}, while the long-range behaviour is related to the zero-point
motion
of plasma oscillations\cite{bp,rene1}. The one-body term
$\chi_{\sigma_i}({\bf
r})$, which is absent in the case of a uniform electron gas, allows a
variational adjustment of the electron density in the presence of the
two-body
term. Despite the apparent simplicity of the Slater-Jastrow trial wave
function of Eq. (\ref{Jastrow}), VMC has proved very successful in
applications
to bulk properties of solids. In particular, VMC methods were used to
calculate cohesive energies of solid
materials\cite{fahy,li,rajagopal,ken}, 
showing that they are typically an order of magnitude more accurate than
cohesive
energies obtained within the HF or LDA approaches.

The main drawback of VMC is that the accuracy of the results
is entirely dependent on the quality of the trial wave function.  
The DMC method overcomes this limitation, by
using a
projection technique to enhance the ground-state component of a
starting trial wave function\cite{Reynolds}. In DMC, the mathematical
equivalence
between the imaginary-time many-electron Schr\"{o}dinger equation and a
real-time diffusion equation is used to solve the former by a simulated
numerical diffusion of a collection of fictitious particles (walkers).
Each walker corresponds to a point in the $3N$ dimensional
space of electron configurations. The density of walkers represents the
many-electron wave function, and the propagation of walkers, according
to the
imaginary-time Schr\"{o}dinger equation, exponentially projects out the
ground
state from the initial state (provided that the initial state is not
orthogonal
to the ground state). In this way, an exact 
numerical representation of the ground state can be obtained 
from a reasonably accurate initial guess. The initial wave function for
the
more accurate and computationally expensive DMC calculation is usually
provided
by VMC methods.

The analogy between the imaginary-time Schr\"{o}dinger equation and the
diffusion equation only works if the  many-body wave function is a
positive (or
negative) function, which can then be interpreted as a density
distribution.
As for fermions the antisymmetric many-electron wave function 
must take both positive and negative values, there is a sign problem in
DMC
calculations for these particles. The common way to circumvent this
problem is
to use the fixed-node approximation\cite{Reynolds,Moskowitz}. Within
this
approach, for a given trial many-electron wave function one defines a
trial
many-electron nodal surface on which the trial function is zero and
across which
it changes sign. The fixed-node DMC algorithm then produces the
lowest-energy
many-electron state with the given nodal surface. The resulting
fixed-node
errors are typically about 5\% of the correlation energy 
(defined as the difference between the {\it exact} and the HF 
ground-state energies) \cite{rmp}. In most systems exchange dominates
the
correlation by almost an order of magnitude, and fixed-node errors are,
therefore, reduced to about $0.5\%$ of the total ground-state energy.
Another possible source for systematic error in QMC calculations is the
use of a finite simulation cell to model an extended
system\footnote{Within a
many-electron formulation, the Schr\"{o}dinger equation cannot be solved
by considering only a single primitive unit cell of the periodic nuclear
potential, since the e-e interaction breaks the lattice symmetry.}. 
We shall return to these finite-size errors in section III.B, where 
we will examine QMC calculations of the jellium surface energy. 

Although VMC and DMC methods are best suited to study the ground state,
they
have been shown to also provide some information about specific excited
states.
This subject has recently been reviewed by Foulkes {\it et
al\,}\cite{rmp}.

\subsection{Green-function theory}

The Kohn-Sham eigenvalues are often found to be in good agreement with
the
quasiparticle energies measured in photoemission experiments.
Nevertheless,
they have no clear physical meaning, except for the highest occupied 
one which
corresponds to the negative of the ionisation energy\cite{vonbarth1}, 
and there are some
serious discrepancies with measured single-particle excitation energies.
Also,
the knowledge of the finite lifetime of excited states is known to 
be outside the domain of DFT.

A proper way of calculating single-particle excitation energies and
lifetimes is
provided by the Green-function theory\cite{Galitskii} and the  {\it
quasiparticle} concept accounting, in a many-electron system, for the
ensemble
of each electron and its own xc hole. 
With $|N,s>$ as some many-body eigenstate
$s$ ($s=0$ for the ground state) of $N$ electrons, one defines the 
quasiparticle 
amplitude $\Psi_s(\br)$ and the quasiparticle energy $E_s$
as\cite{Hedin}
\begin{equation}
\Psi_s(\br)=<N,0|\hat\psi(\br)|N+1,s>,\qquad
E_s=E_{N+1,0}-E_{N,0}\quad{\rm
for}\quad E_s\ge\mu
\end{equation}
and
\begin{equation}
\Psi_s(\br)=<N-1,s|\hat\psi(\br)|N,0>,\qquad
E_s=E_{N,0}-E_{N-1,s}\quad{\rm
for}\quad E_s<\mu,
\end{equation}
where $\hat\psi(\br)$ is a field operator in the Heisenberg
representation that destroys an electron at point $\br$, and $\mu$
represents
the chemical potential.

The one-particle Green function\cite{Galitskii} determines the energy
and the
damping of the quasiparticles. This can be shown explicitly by
introducing the
complete set of many-body eigenstates of the full Hamiltonian for the
$N+1$ or
$N-1$ particles in the definition of the one-particle Green
function. One obtains
\begin{equation}\label{gf}
G(\br,\br';t)=-i\sum_s\Psi_s(\br)\,\Psi_s^*(\br)\,{\rm e}^{-i\,E_st},
\end{equation}
which identifies $-2\,{\rm Im}E_s$ as the inverse quasiparticle
lifetime.
Furthermore, Fourier transformation of Eq. (\ref{gf}) to energy space
yields
\begin{equation}\label{spectral}
G(\br,\br';E)=\sum_s{\Psi_s(\br)\Psi_s^*(\br')\over E-E_s},
\end{equation}
which is known to satisfy the Dyson equation
\begin{equation}\label{Dyson}
G(\br,\br';E)=G^0(\br,\br';E)+\int d\br_1\int
d\br_2\,G^0(\br,\br_1;E)\,\Sigma(\br_1,\br_2;E)\,G(\br_2,\br';E).
\end{equation}
Here $G^0(\br,\br';E)$ represents the non-interacting one-particle Green
function, and $\Sigma(\br,\br';E)$ is the so-called self-energy of the
quasiparticle, which is a nonlocal, energy-dependent, non-Hermitian
operator
accounting for all xc effects beyond the Hartree approximation. By
inserting the
spectral representation of Eq. (\ref{spectral}) into Eq. (\ref{Dyson}),
one finds
the quasiparticle equation
\begin{equation}\label{quasi}
\left[-{1\over 2}\nabla^2+V_{ext}(\br)+\int d{\bf
r}'\,\frac{n({\bf
r}')}{|{\bf r}-{\bf r}'|}\right]\Psi_s(\br)+
\int d\br'\,\Sigma(\br,\br';E_s)\,\Psi_s(\br')=E_s\,\Psi_s(\br).
\end{equation}

For a system of interacting electrons, there is little hope in solving
the
quasiparticle equation exactly. Hence, one usually resorts to
perturbation
theory and chooses, as a starting point, a suitable single-particle
Hamiltonian $\hat H_0$ whose wave functions $\phi_s(\br)$ and energies
$\varepsilon_s$ approach the quasiparticle wave functions $\Psi_s(\br)$
and
energies $E_s$. In usual practice, one takes the LDA Kohn-Sham
Hamiltonian
\begin{equation}\label{lda}
\hat H_0=-\frac{1}{2}\nabla^2+V_{\rm ext}({\bf r})+\int d{\bf
r}'\,\frac{n({\bf
r}')}{|{\bf r}-{\bf r}'|}+V_{xc}^{LDA}({\bf r}),
\end{equation}
where
\begin{equation}
V_{xc}^{LDA}(\br)=\left.{d\left[n\,\varepsilon_{xc}^{unif}(n)\right]\over
dn}\right|_{n=n(\br)}.
\end{equation}
The quasiparticle energy is then determined from
\begin{equation}\label{quasienergy}
E_s\approx\varepsilon_s+Z_s\,\Delta\Sigma_s(\varepsilon_s),
\end{equation}
where
\begin{equation}
\Delta\Sigma_s(E)=\int d\br\int
d\br'\,\phi_s^*(\br)\left[\Sigma(\br,\br';E)-
V_{xc}(\br)\delta(\br-\br')\right]\phi_s(\br'),
\end{equation}
and
\begin{equation}
Z_s=\left[1-\left.{\partial
\Delta\Sigma_s(E)\over\partial E}\right|_{E=\varepsilon_s}\right]^{-1}
\end{equation}
is the so-called quasiparticle weight or renormalisation factor. On the
{\it energy-shell}, this factor is taken to be unity.

The exact self-energy can be obtained, in principle, from an iterative
solution
of Hedin's equations\cite{Hedin} in combination with the Dyson equation.
However, to obtain explicit results one usually resorts to an expansion
in
powers of the time-ordered screened interaction $W(\br,\br';E)$. The
leading-order term of this expansion is the so-called GW approximation:
\begin{equation}\label{gw}
\Sigma(\br,\br';E)=i\int_{-\infty}^\infty {dE'\over 2\pi}\,e^{-i\eta
E'}\,
G(\br,\br';E-E')\,W(\br,\br';E'),
\end{equation}
which can also be obtained as the first iteration of Hedin's equations
by
simply
neglecting vertex corrections. The screened interaction can be expressed
in
terms of the density-response function $\chi(\br,\br';E)$, as follows
\begin{equation}\label{w}
W({\bf r},{\bf r}';E)=v({\bf r}-{\bf r}')+\int{\rm d}{{\bf
r}_1}\int{\rm d}{{\bf r}_2}\,v({\bf r}-{\bf r}_1)\,
\chi({\bf r}_1,{\bf r}_2,E)\,v({\bf r}_2-{\bf r}'),
\end{equation}
where $v({\bf r}-{\bf r}')$ represents the bare Coulomb interaction. If
one replaces the screened interaction by $v({\bf r}-{\bf r}')$ the
self-energy of
Eq. (\ref{gw}) reduces to the Hartree-Fock self-energy, which sometimes
leads
to unphysical results, especially in the case of metals where screening
plays a crucial role.

Most current GW calculations simply replace the
{\it exact} one-particle Green function entering Eq. (\ref{gw}) by the
non-interacting Green function $G^0(\br,\br';E)$ pertaining to the LDA
Kohn-Sham Hamiltonian of Eq. (\ref{lda}). On the same level of approximation and
neglecting all vertex corrections, the screened interaction entering Eq.
(\ref{gw}) is usually obtained from Eq. (\ref{w}) with the density-response
function evaluated in the random-phase approximation (RPA)\footnote{This
DFT-based RPA differs from the less realistic {\it actual} RPA, which is
determined from the non-interacting Green function associated with the Hartree
[$V_{xc}^{LDA}(\br)=0$ in Eq. (\ref{lda})] Hamiltonian.}:     
\begin{equation}\label{rpa}
\chi({\bf r},{\bf r}';E)=\chi^0({\bf r},{\bf r}';E)
+\int{\rm d}{\bf r}_1\int{\rm d}{\bf r}_2\,\chi^0({\bf r},{\bf
r}_1;E)\,v({\bf r}_1-{\bf r}_2)\,
\chi({\bf r}_2,{\bf r}';E),
\end{equation}
$\chi^0({\bf r},{\bf r}';E)$ being the non-interacting density-response
function
\begin{equation}\label{g0}
\chi^0({\bf r},{\bf r}';E)=-2\,i\int dE'\,G^0(\br,\br';E')\,G^0(\br,\br';E+E').
\end{equation}

A survey of the theory underlying the GW formalism and its applications
can be
found in Refs.\onlinecite{gw} and \onlinecite{Farid}. For improvements
of the
{\it standard} ($G^0W^0$) quasiparticle calculations, including
self-consistency and beyond-GW approximations, we refer the reader to
the recent
review by Aulbur, J\"onsson, and Wilkins\cite{gw2}.

\section{Ground-state properties of surfaces}
\label{sec:prog-ground}

The simplest model to investigate the many-body properties of simple
metal surfaces is the well-known {\it jellium} model. 
Within this model, valence electrons are described by an 
inhomogeneous assembly of free electrons moving 
in a uniform neutralising positive background, which is abruptly 
truncated at a plane. In this section, we
focus on a description of many-body properties of jellium surfaces,
and we also survey, when available, more realistic calculations where
the effect of the discrete crystal lattice is taken into account.
In the discussion of the jellium surface, the $z$ axis is taken to be
normal to
the surface, with the metal occupying the $z>0$ half-space. The
positive-background charge density $\bar {n}$ is expressed in terms of
the Wigner
radius
$r_s$, as
$1/\bar{n}=(4\pi/3)r_s^3$.

\subsection{Pair-distribution function and exchange-correlation hole}

Central quantities in the discussion of xc effects in an interacting
many-electron system are the pair-distribution function $g(\br,\br')$
and the xc-hole charge density $n_{xc}(\br,\br')$.
The pair-distribution 
function is defined as the probability of finding an electron 
at point $\br'$  if there is already one electron at 
$\br$
\begin{equation}\label{wave}
g(\br,\br')={1\over
n(\br)n(\br')}\left[<N,0|\hat n(\br)\,\hat n(\br')
|N,0>-n(\br)\delta(\br-\br')\right],
\end{equation}
where $|N,0>$ represents the many-body ground state of $N$ interacting
electrons, $n(\br)$ is the electron density, and $\hat n(\br)$
represents the electron-density operator. 
Making use of the zero-temperature fluctuation-dissipation
theorem\cite{Callen},
and  by a suitable choice of integration contours\cite{Pines},
the pair correlation function can be written in terms of 
the retarded density-response function $\chi({\bf r},{\bf r}';E)$ 
of the many-electron system\footnote{For positive
energies this function coincides with its time-ordered counterpart,
which enters the self-energy formalism described in the preceding
section.} 
\begin{equation}\label{chig}
g(\br,\br')=1+{1\over n(\br)n(\br')}\left[-{1\over\pi}\int_0^\infty
dE\,\chi(\br,\br';i\,E)- n(\br)\,\delta(\br-\br')\right].
\end{equation}

The total interaction energy $E_{ee}$ of the many-electron system is
obtained
as follows
\begin{equation}
E_{ee}=\frac{1}{2}\int d\br\int d\br'\,\frac{n(\br)g(\br,\br')n(\br')}
{|\br-\br'|},
\end{equation}
which can be written in the form
\begin{equation}
E_{ee}=\int d\br\,n(\br)\,\varepsilon_{ee}(\br;[n(\tilde\br)]),
\end{equation}
$\varepsilon_{ee}(\br;[n(\tilde\br)])$ representing the interaction energy
per
particle at point $\br$. This interaction energy can be separated
into
electrostatic (Hartree) and xc terms,
\begin{equation}\label{ee}
\varepsilon_{ee}(\br;[n(\tilde\br)])=
\frac{1}{2}\int d\br'\,\frac{n(\br')}
{|\br-\br'|}+\frac{1}{2}\int 
d\br'\,\frac{n_{xc}(\br,\br')}{|\br-\br'|},
\end{equation}
$n_{xc}(\br,\br')$ being the so-called xc-hole charge density
\begin{equation}\label{hole}
n_{xc}(\br,\br')=n(\br')\left[1-g(\br,\br')\right].
\end{equation}
This quantity represents the depletion of the average electron density
$n(\br')$ due to the presence of an electron at $\br$, which is dictated
by the
combined effect of the Pauli exclusion principle and Coulomb
correlations, and satisfies the sum-rule
\begin{equation}
\int d\br'\,n_{xc}(\br,\br')=1.
\end{equation}

Eqs. (\ref{wave}) and (\ref{hole}) represented the starting point
for recent QMC calculations of $n_{xc}$ in bulk silicon\cite{Hood}, the jellium
surface\cite{dmc,acioli}, and strongly inhomogeneous model
solids\cite{Nekovee3}.  Alternatively, Eq. (\ref{chig})
allows a calculation of the pair-distribution function and
the xc hole from the knowledge of the density-response function of the
system. 
In a DFT-based RPA, this function is usually obtained from Eq.
(\ref{rpa}) with
the use of the single-particle eigenfunctions and eigenvalues
of the LDA Kohn-Sham Hamiltonian of Eq. (\ref{lda}). In the framework of
time-dependent density-functional theory\cite{tddft,peter}, the {\it
exact}
density-response function satisfies the integral equation
\begin{equation}\label{chi}
\chi({\bf r},{\bf r}';E)=
\chi^0({\bf r},{\bf r}';E)+\int d{\bf r}_1\int{\rm
d}{\bf r}_2\,\chi^0({\bf r},{\bf
r}_1;E)\left[v({\bf r}_1,{\bf r}_2)
+f_{xc}({\bf r}_1,{\bf r}_2;E)\right]\chi({\bf r}_2,{\bf
r}';E),
\end{equation}
where
\begin{equation}\label{kernel}
f_{xc}({\bf r},{\bf r}';E)={\delta v_{xc}\left[n({\bf
r},E)\right]\over\delta n({\bf r}',E)},
\end{equation}
$v_{xc}\left[n({\bf r},E)\right]$ representing the exact energy-dependent xc
potential, and $\chi^0({\bf r},{\bf r}';E)$ being the
density-response function of non-interacting Kohn-Sham electrons.
In the so-called time-dependent local-density approximation
(TDLDA)\cite{tdlda} or, equivalently, adiabatic local-density
approximation (ALDA), the Kohn-Sham effective potential is obtained in the LDA
and the {\it exact} xc kernel of Eq. (\ref{kernel}) is replaced, for all
energies, by
\begin{equation}\label{kernellda}
f_{xc}^{ALDA}({\bf r},{\bf
r}';E)=\left.{d^2\left[n\,\varepsilon_{xc}^{unif}(n)\right]\over
dn^2}\right|_{n=n(\br)}\delta(\br-\br').
\end{equation}

Assuming, within a jellium model of the surface, translational
invariance
in the plane normal to the $z$ axis, the xc-hole charge density is found
to
be given by the following expression:
\begin{equation}\label{ens}
n_{xc}(r;z,z')={1\over 2\pi n(z)}\int
dq_\parallel\,q_\parallel\,J_0(q_\parallel
r)\left[{1\over\pi}\int_0^\infty
dE\,\chi(z,z';q_\parallel,i\,E)]+n(z)\delta(z-z')\right],
\end{equation}
which depends on the distance $r=|\br_\parallel-\br_\parallel'|$
projected in
the plane of the surface. If the actual density-response function
$\chi(z,z';q_\parallel,i\,E)$ is replaced by its non-interacting
counterpart $\chi^0(z,z';q_\parallel,i\,E)$, then Eq. (\ref{ens}) yields
the exchange-only or Fermi hole\footnote{If $\chi^0(z,z';q_\parallel,i\,E)$
were determined from the eigenfunctions and eigenvalues of the non-local
Hartree-Fock Hamiltonian, this Fermi hole would coincide with that obtained in
the Hartree-Fock approximation.}.

In the simplest possible microscopic model of the jellium surface,
namely the so-called infinite barrier model (IBM), the one-electron wave
functions describing motion normal to the surface are simply sine
functions.
Using  these wave functions, the exchange hole was studied by
Juretschke\cite{Jur} and others\cite{Moore}, and the xc hole was later
investigated in the RPA by Inglesfield and Moore\cite{Ing1}. However,
the IBM
does not permit the electron density to relax beyond the infinitely high
potential barrier outside the surface, and yields a poor representation
of both
the electron density and the xc hole.

The  behaviour of the exchange hole for an electron that is located well
outside
the surface was investigated by Sahni and Bohnen\cite{Sahni1} with
the use of one-electron wave functions that were generated from a
linear-potential model. These authors found  that inside the solid, and
up to
the position of the jellium edge, the x-hole density $n_x(r;z,z')$
shows a behaviour similar to that obtained within the IBM. As the
electron overcomes the jellium edge, the exchange hole was found to
remain
behind and to be localised near the surface region, as within the IBM.
However,
when the electron is removed further away from the surface into the
vacuum the
exchange hole, instead of staying localised at the surface, was found to
spread
throughout the entire solid.  This behaviour was further confirmed by
the
calculations reported by Harbola and Sahni\cite{Sahni2}, who used a
step-potential model of the jellium surface. These authors also
considered the
x-hole charge distribution in the plane parallel to the surface, and
found that
it differs significantly from the classical image charge distribution.

Recently, preliminary self-consistent calculations of both exchange
and correlation contributions to the xc-hole charge density at a jellium
surface were reported\cite{Pitarke0}, as obtained from Eqs. (\ref{chig})
and
(\ref{hole}) with a full description of the DFT-based RPA
density-response
function of a jellium slab. These calculations showed that for an
electron well
outside the surface the x-only charge density displays several maxima
within the
solid, spaced at intervals of half a Fermi wavelength, in agreement with
the
results of Sahni and Bohnen, and confirmed the interesting, albeit
expected
result, that the build up of the image charge is entirely due to the
surface-localised part of the correlation hole.

Fig. 1 shows contour plots of the fully self-consistent DFT-based RPA xc
hole
near a jellium surface with the positive-background charge density $\bar
n$ of
aluminium ($r_s=2.07$). In the interior of the
metal the xc hole is a sphere cantered at the position of the electron,
as in
the case of a uniform electron gas. As the electron moves towards the
jellium
edge, the xc-hole charge density $n_{xc}$ begins to distort and starts
to
lag behind. Finally, as the electron is removed far outside the
surface the xc hole flattens and remains localised  near the surface,
becoming
an image-charge distribution located at the so-called image plane.

The issue of the physical origin of the image potential has been the
source of considerable controversy over the years. The asymptotic
behaviour of
the xc potential $V_{xc}(\br)$ of DFT at large distances outside a metal
surface
was examined by Sham\cite{Sham1} and by Eguiluz {\it et
al.}\cite{Eguiluz1}
through the following exact integral equation, which relates
$V_{xc}(\br)$ to the xc self-energy
$\Sigma_{xc}(\br,\br';E)$\cite{Sham0}:
\begin{equation}\label{integral}
\int d\br'\,V_{xc}(\br')\int dE\,G^0(\br,\br';E)\,G(\br',\br;E)=
\int d\br_1\int d\br_2\int
dE\,G^0(\br,\br_1;E)\,\Sigma_{xc}(\br_1,\br_2;E)
\,G(\br_2,\br';E).
\end{equation}
Both Sham\cite{Sham1} and Eguiluz {\it et al.}\cite{Eguiluz1} concluded
that
the exchange self-energy $\Sigma_x(\br,\br';E)$ yields a $1/z^2$
behaviour for
large $z$, while the correlation self-energy yields the image-like $1/z$
limit,
thereby confirming the long-standing believe that the image-potential
structure is a pure Coulomb-correlation effect.

Harbola and Sahni suggested an alternative procedure for constructing
the xc
potential that enters the Kohn-Sham equations of DFT\cite{Harbola}.
These authors
interpreted $V_{xc}(\br)$ as the work required to remove an electron
from the
solid against the electric field of its xc-hole charge distribution, and
showed
that the exchange-only potential $V_x(\br)$ merges with the {\it
classical} image
potential. Although this conclusion seemed to be in disagreement with
the
numerical jellium-slab $G^0W^0$ calculations reported in
Ref.\onlinecite{Eguiluz1}, it was then argued that while
correlation gives the asymptotic image limit for slab geometries
exchange does
provide the limit in the case of the semi-infinite solid\cite{Sahni3}.
Nevertheless, the conclusion of Harbola and Sahni\cite{Harbola} is still
in
disagreement with the earlier investigations of Ref.\onlinecite{Sham1},
and this
issue deserves further investigation. Alternatively, the image potential
felt by
electrons well outside the surface can be obtained from the asymptotic
behaviour
of the interaction energy per particle
$\varepsilon_{ee}(\br;[n(\tilde\br)])$
of Eq. (\ref{ee}), which is dictated by the interaction between each
electron
and its own xc hole. Work along these lines is now in
progress\cite{Pitarke2}.

\subsection{Surface energy}

One of the surface properties most sensitive to many-body effects is the
surface energy. This static physical quantity is defined as the energy
per unit area required to split the solid into two separate halves along
a plane,
i.e., 
\be
\sigma=\frac{2\,E[n(\br)]-E[n'(\br)]}{2\,A},
\ee
where $n(\br)$ represents the ground-state electron-density distribution
for each
half of the solid after it is split, and $n'(\br)$ is that for the
unsplit
solid. Following the DFT energy-functional partitioning of Eq.
(\ref{xc}), the
surface energy can be written as the sum of three terms:
\begin{equation}
\sigma=\sigma_s+\sigma_{es}+\sigma_{xc},
\end{equation}
where $\sigma_s$ represents the kinetic surface energy of
non-interacting
Kohn-Sham electrons, $\sigma_{es}$ is the electrostatic surface energy
due to
all positive and negative charge distributions in the solid, and
$\sigma_{xc}$ is the xc contribution to the surface energy. For a
jellium
surface, one writes
\begin{equation}\label{eqxc}
\sigma_{xc}=\int_0^\infty
dz\,n(z)\,
\left\{\varepsilon_{xc}(z;[n(\tilde
z)])-\varepsilon_{xc}^{unif}(\bar{n})\right\},
\end{equation}
where 
$\varepsilon_{xc}(z;[n(\tilde z)])$ is the xc energy per 
particle at point $z$ and $\varepsilon_{xc}^{unif}(\bar{n})$ is the xc
energy
per particle of a uniform electron gas of density $\bar n$. In the LDA,
$\varepsilon_{xc}(z;[n(\tilde z)])$ is simply replaced
by $\varepsilon_{xc}^{unif}(n(z))$.

The first self-consistent DFT calculation of the jellium surface energy
was
carried out by Lang and Kohn\cite{Lang}, in the LDA approximation. Lang
and Kohn
also discussed the effect of the crystal lattice, within first-order
perturbation
theory, and this work was later generalised to treat the
discrete-lattice
perturbation variationally\cite{Monnier}. Nowadays, the
full inclusion, within the LDA, of lattice effects poses no
difficulties;
however, the question of the impact of nonlocal xc effects on the
surface energy,
and their interplay with the strong charge  inhomogeneity at the
surface, has
remained unsettled over the years.

The original discussion about the effect of nonlocal Coulomb
correlations was centred around 
the surface-plasmon contribution to the surface energy.
Craig\cite{Craig}
and Schmidt and Lucas\cite{Lucas} independently claimed that the
surface energy of simple metals can be accounted for by the change of
zero-point
energy of the plasma oscillations resulting from the cleavage of the
metal,
which is a nonlocal
effect not contained in the LDA calculation of Lang and Kohn. 
However, the actual
magnitude of this contribution became a matter of
controversy\cite{Feibelman},
and more careful calculations were then presented. The xc surface
energy was first obtained in
the RPA for a free-electron gas that was terminated abruptly 
at a plane\cite{Harris,Jonson}, and more
elaborate RPA calculations were then reported for the case in which the
surface potential is taken to be an infinite barrier\cite{Ingel2}. A
comparison
between these
IBM-based RPA calculations and a local-density calculation of
$\sigma_{xc}$ for
the IBM density profile yielded a difference of $\sim 10\%$, though much
larger
differences were found for the separate exchange and correlation
contributions to $\sigma_{xc}$\cite{Langreth1}. Lang and Sham then
argued that a
regular power series in density gradients does not exist for separate
exchange
and correlation energies, and showed that the LDA is satisfactory for
the
combined xc term but not for exchange and for correlation
separately\cite{ls}.

RPA and ALDA calculations of jellium surface energies that are based on
a fully self-consistent description of the one-particle Hartree or LDA
orbitals
have been performed only very recently\cite{Pitarke1,Pitarke3}. In these
calculations, both $\varepsilon_{xc}(z;[n(\tilde z)])$ and
$\varepsilon_{xc}^{unif}(\bar{n})$ were obtained from a
coupling-constant
integration of the xc contribution to the interaction energy of Eq.
(\ref{ee})\cite{Gunnarsson,Perdewl},
\begin{equation}
\varepsilon_{xc}(\br;[n(\tilde\br)])={1\over 2}\,\int_0^1d\lambda\,\int 
d\br'\,{n_{xc}^\lambda(\br,\br')\over|\br-\br'|},
\end{equation}
where $n_{xc}^\lambda(\br,\br')$ is the xc-hole charge density of a
fictitious
system so chosen that as the e-e interaction strength $\lambda$
is varied in the interval $[0,1]$ the corresponding density equals its
fully
interacting value, i.e., $n_\lambda(\br)=n(\br)$. In
Refs.\onlinecite{Pitarke1}
and \onlinecite{Pitarke3}, LDA surface energies were also computed, as
obtained
from the RPA and ALDA xc energies per particle of a uniform electron
gas, and a comparison of local versus nonlocal surface energies
supported the
conclusion that the error introduced by the local-density
approximation is small.

The RPA is exact for exchange and long-range correlations, but it is
known to be a poor approximation for short-range correlations. 
Hence, Kurth and Perdew\cite{Kurth1} constructed a GGA for the
short-range
correlation energy in combination with the DFT-based RPA calculations of
Refs.\onlinecite{Pitarke1} and\onlinecite{Pitarke3} for long-range
correlation,
and obtained xc surface energies that are larger than the LDA prediction
by about
$2\%$. Furthermore, it was argued that while the RPA is not a good
approximation for the total correlation energy, it seems to be a
surprisingly good approximation for certain changes in the correlation
energy,
such as those arising in surface energies\cite{Kurth2}.

Two new approaches have been developed recently, which yield surface
energies agreeing to within $1\%$ with the so-called RPA+ calculations
of 
Ref.\onlinecite{Kurth1}. These are: (a) a meta-GGA\cite{meta}, and (b) a
new
wave-vector interpolation as a long-range correction to the GGA exchange
and
correlation\cite{Kurth3}. Nonetheless, these calculations strongly
disagree with
the result of wave-function-based Fermi-hypernetted-chain
(FHNC)\cite{fhnc} and
DMC\cite{dmc,acioli} calculations, which predict surface energies that
are
significantly higher than those obtained in the local-density
approximation.

Acioli and Ceperley\cite{acioli} performed fixed-node DMC calculations
of the
surface energy for a jellium slab that is contained in a finite
simulation cell.
They obtained surface energies that are systematically larger than the
corresponding LDA results. Although for a high-density metal
($r_s=2.07$) the
DMC surface energy was found to be close to the LDA prediction, the
difference with the LDA surface energy 
was found to be as large as $50\%$ for $r_s=4$, in
agreement with the FHNC calculations of Krotscheck and Kohn\cite{fhnc}
(the absolute difference between DMC surface energies and those obtained
in the LDA was found
to be of about 150 $erg/cm^2$ for all densities under study, i.e.,
$2.07\leq r_s\leq 4$). The discrepancy between DFT and
wave-function-based
surface energies is surprising: while fixed-node DMC calculations are
often regarded as essentially exact, all the current DFT-based estimates
of the jellium surface energy\cite{meta,Pitarke1,Pitarke3,Kurth1,Kurth2,Kurth3}
seem to be consistent with the conclusion that LDA and GGA calculations for
real surfaces provide energies at least as accurate as those derived from the
experiment\cite{methfessel,Skriver,Vitos}.

A possible source of systematic error in QMC calculations for extended
systems
is the use of a finite simulation cell. Hence, one needs to extrapolate
the
ground-state energy $E_\infty$ of the extended medium from that obtained
for
the $N$-particle system, $E_N$. In the case of a uniform system, one 
usually uses the extrapolation formula    
\be 
\label{eq:finsize}
E_\infty=E_N+a\,(T_\infty-T_N)+\frac{b}{N},
\ee
where the parameters $a$ and $b$ are fitted by considering the energy 
of increasingly larger systems. $T_\infty$ and $T_N$ represent the
kinetic
energies of the non-interacting infinite and finite systems, and the
$b/N$ term accounts for the so-called Coulomb finite-size effects
arising from
the e-e interaction-energy term.
The non-interacting term $(T_\infty-T_N)$ is generally obtained with the
use of
one-particle LDA calculations, 
and the parameter $a$ ($a\sim 1$) accounts for the
difference between the kinetic energies of the interacting and the
non-interacting systems. For non-uniform systems, there are
also finite-size errors due to the electrostatic-energy
contribution to the ground-state energy, which only depend on the
electron
density. As the LDA densities are found to be rather accurate, these finite-size
errors can
also be estimated within the LDA. 

Acioli and Ceperly\cite{acioli} estimated the finite-size corrections to
their DMC surface energies from the difference between the LDA surface
energies
of their finite simulation cell and a jellium slab that is infinitely 
extended in the plane parallel to the surface. 
Although this estimate accounts for kinetic and (part of) electrostatic
contributions to the finite-size effects, there are additional
(Coulomb) finite-size errors, not encountered in the LDA, which are originated
in the periodic Ewald interaction used to model the e-e Coulomb
interaction in a periodic geometry.

In order to estimate the magnitude of the Coulomb finite-size 
corrections, which are not included in the DMC surface energies of
Ref.\onlinecite{acioli}, we used the VMC method to calculate the surface
energy
of a jellium slab with $r_s=3.25$. For the trial wave function we used
the
standard Slater-Jastrow form of Eq. (\ref{Jastrow}) in combination with 
the
one-body term $\chi_i(\br)$ due to Fahy {\it et al.}\cite{maltesa}. The
calculations were performed with the use of a faced-cantered cubic (fcc)
supercell containing $N=284$ spin-unpolarised electrons. We modelled the
e-e
interaction energy using either the standard Ewald interaction, as used
by
Acioli and Ceperley\cite{acioli}, or the model periodic Coulomb (MPC)
interaction of Ref.\onlinecite{mpc}, which virtually eliminates Coulomb
finite-size effects. Our calculations indicate that for $N=284$ and
$r_s=3.25$ Coulomb finite-size corrections to the VMC simulations are
positive
and amount to $3.9\%$ of the surface energy.  Since these corrections
decay as
$1/N$ with the size of the system, we estimate that Coulomb 
finite-size corrections to the DMC calculations of 
Acioli and Ceperley amount to
$\sim 6\%$ of the jellium surface energy for $r_s=3.25$.
Our preliminary VMC calculations yield surface energies of
$433$ and $450\,{\rm erg/cm^2}$, respectively, depending on whether the
Ewald or the MPC interaction is employed.
These surface energies are higher than
predicted by the DMC calculations of Acioli and Ceperley, which yield
for
$r_s=3.25$ a surface energy of $360\,{\rm erg/cm^2}$, and are well above the
LDA
prediction of $220\,{\rm erg/cm^2}$ for this electron density\footnote{In these
calculations we used the DMC total energy of as quoted by 
Acioli and Ceperley for the total energy of the uniform electron gas.
Using our own VMC energy of a uniform electron gas calculated for 
a system with $N=284$ electrons yields 
$\sigma=338\,{\rm erg/cm^2}$, which still is much larger
than the LDA surface energy.}.

Hence, finite-size corrections do not seem to account for the existing
large
differences between QMC and LDA surface energies, which are not
present in DFT-based calculations, and a satisfactory understanding of
nonlocal correlation effects in the surface energy of simple metals
remains to be
given in the future. Although the LDA and GGA for {\it real} metals have
been
found to provide surface energies in fair agreement with the experiment,
this
might well be due to a cancellation between the errors introduced by the
LDA in the total energy of both the infinite and the semi-infinite solid.
Such a cancellation of errors could not occur in the case of jellium
surface energies, as the LDA is exact for the infinitely extended jellium.

\subsection{Work function}

The work function $\Phi$ is the minimum energy required 
to remove an electron from the solid to a macroscopic distance 
outside the surface. In a many-body context (and at $T=0$) 
this corresponds to a process in which a many-body system composed of
$N$
electrons is brought from its ground state to an excited state with
$N-1$ 
electrons in the lowest possible state in the metal and one electron at
rest at infinity. Hence, one writes
\be\label{work}
\Phi=E_{N'}-E_N,
\ee
where $E_N$ is the ground-state energy of the semi-infinite 
solid and $E_{N'}$ is the energy of the excited state.
The long-range image potential tail of the surface-electron
interaction potential ensures that the electron remains bounded to the
surface, occupying a bound state $n$ ($n=1$ for the lowest-lying state) of the
image potential with energy\cite{Echenique}
\be\label{image}
\epsilon_n=\phi_\infty-\frac{1}{4(n+a)^2},
\ee
where $a$ is the so-called quantum defect, and $\phi_\infty$ is 
the electrostatic potential well outside the
surface. Thus, the minimum energy of 
an electron moving freely at a macroscopic distance outside the surface
is
\begin{eqnarray}
\epsilon&=&\lim_{n\to\infty}\left[\phi_\infty-\frac{1}{4(n+a)^2}\right]\cr\cr
&=&\phi_\infty,
\end{eqnarray}
and the total energy of the lowest-lying excited state of the
semi-infinite
solid with one electron at infinity is
\begin{equation}
E_N'=E_{N-1}+\phi_\infty.
\end{equation}

As the highest occupied DFT eigenvalue $E_F$ equals the (negative of)
the
ionisation energy
$E_N-E_{N-1}$, Eq. (\ref{work}) may be written in the equivalent
one-body form
\be 
\label{eq:workfun3}
\Phi=\phi_\infty-E_F,
\ee
or, equivalently,
\begin{equation}\label{real}
\Phi=\Delta\phi-\bar E_F,
\end{equation}
where $\Delta\phi$ is the so-called surface dipole barrier, and $\bar
E_F$
represents the Fermi level relative to the mean interior electrostatic
potential.

For a jellium surface\cite{Lang}, 
\be\label{bar}
\bar E_F=\frac{1}{2}k_F^2+\mu_{xc}(\bar{n}),
\ee
where $\mu_{xc}(\bar{n})$ represents the xc potential of 
the uniform electron gas, and $k_F=(3\pi^2\bar{n})^{1/3}$ is the Fermi
momentum.
Using Eq. (\ref{bar}) one then obtains an exact general expression for
the
jellium work function:
\be
\Phi_{J}=\Delta\phi-\frac{1}{2}k_F^2-\mu_{xc}(\bar{n}).
\ee
Both $k_F$ and $\mu_{xc}(\bar{n})$ are known properties of
the uniform
electron gas and one only needs, therefore, to approximate the dipole
barrier, which is usually evaluated in the LDA.

LDA self-consistent calculations of the work function of a jellium
surface
were carried out by Lang and Kohn\cite{Lang2}. The LDA xc potential
decays
exponentially
outside the jellium surface and does not reproduce, therefore, the
correct
image-like asymptotic behaviour. Hence, LDA orbitals yield a too rapidly
varying electron-density profile and values of the dipole barrier and
the work
function that are too high. Nonlocal calculations of the work function
have
been performed with the use of wave-function-based FHNC\cite{fhnc} and
DMC\cite{acioli} methods, and
they indicate that the actual work function of jellium surfaces lies
about
$0.3-0.5\,{\rm eV}$ lower than their LDA counterparts. Nevertheless, an
accurate determination of DMC work functions still requires longer runs
and
thicker simulation slabs, in order to eliminate the statistical
fluctuations in
the electron-density profiles and to avoid the presence of
considerable finite-size effects.

In the case of real surfaces the work function is obtained from Eq.
(\ref{real}), and one needs to determine both the dipole barrier and the
Fermi
level $\bar E_F$. Now the bulk properties are not those of a uniform
electron
gas and $\bar E_F$ also needs to be approximated. In actual
first-principles calculations of the work function both $\Delta\phi$ and
$\bar E_F$ are usually obtained within the LDA. As the LDA
uncertainties introduced in the evaluation of $\bar E_F$ tend to cancel out
the corresponding uncertainties introduced in the evaluation of the
dipole barrier, the LDA is expected to perform better in the evaluation
of the work function of real solids than in the case of jellium. Indeed,
LDA work functions of a variety of real solids have been shown to be in
good agreement with experiment. This trend is particularly striking in
the case of $4d$ metals, where the relative difference between 
first-principles LDA and experimental work functions is found to be
within $5\%$. This is shown in Table~\ref{table:workfun}.

%\begin{table}[h]
%\begin{tabular}{lccc}
%Surface   & LDA$^a$  & EXP$^b$   \\ \hline
%Nb(110)   & 4.66 & 4.80 &     \\
%Mo(110)   & 4.94 & 4.95 &     \\
%Pd(111)   & 5.53 & 5.60 &     \\
%AG(111)   & 4.67 & 4.74 &      \\
%Ag(100)   & 4.43 & 4.64 &      \\ 
%Ag(110)   & 4.23 & 4.40 &      \\
%\end{tabular}
%\caption{LDA and experimental values of the work function for low-index
%crystal
%faces of $4d$ metals, taken from $^a\!\!$
%Ref.\protect\cite{methfessel}\protect\- and $^b\!\!$
%Ref.\protect\cite{workfun_exp2}\protect, respectively}
%\label{table:workfun}
%\end{table}

\section{Quasiparticle calculations}
\label{sec:excited}

\subsection{The surface barrier}

The form of the surface barrier outside a solid is of great importance
for the interpretation of a variety of surface-sensitive experiments,
such as
low-energy electron diffraction (LEED)\cite{leed}, scanning tunnelling
microscopy
(STM)\cite{stm,stm2}, and inverse and two-photon photoemission
spectroscopy\cite{Smith,Fauster}. This issue was addressed in
Refs.\onlinecite{Sham1,Eguiluz1,Harbola,Sahni3} through the exact xc
potential
of DFT, i.e., the xc potential felt by Kohn-Sham electrons. A way of
calculating the actual surface barrier felt by quasiparticle states of
the
many-body system is provided by the Green-function theory and, in
particular,
the nonlocal state-dependent electron self-energy $\Sigma(\br,\br';E)$.

Deisz {\it et al.}\cite{Deisz} and White {\it et al.}\cite{Godby}
simulated the
physics of the self-energy near the surface in terms of an {\it
effective}
state-dependent local potential, whose real part was interpreted as the
actual
potential felt by a given quasiparticle state. Comparing the nonlocal
Hamiltonian entering Eq. (\ref{quasi}) with an effective local
Hamiltonian,
this state-dependent effective local xc potential was defined by 
\begin{equation}\label{effective}
V_{loc}(\br)\,\Psi_s(\br)=\int
d\br'\,\Sigma(\br,\br',E_s)\,\Psi_s(\br').
\end{equation}

Calculations of the local potential of Eq. (\ref{effective}) were
reported in Ref.\onlinecite{Deisz} for the lowest-lying image resonance
in a
jellium surface with $r_s=2.07$. This effective potential was found not
to depend
sensitively on the quasiparticle state and to nearly coincide with the
xc
potential of DFT obtained in Ref.\onlinecite{Eguiluz1} by solving an
exact
integral equation, Eq. (\ref{integral}). Nevertheless, the calculations
reported
in Ref.\onlinecite{Godby} for quasiparticle states at the (111) surface
of real
Al, within $1.5\,{\rm eV}$ of the vacuum energy, indicated that the
relative
contributions of exchange and correlation to the image potential felt by
unoccupied quasiparticle states are significantly different from the
corresponding contributions to the exact $V_{xc}$ of DFT. These authors
showed
that in the case of quasiparticle states above the Fermi level the
exchange part
of the effective xc potential of Eq. (\ref{effective}) displays
exponential
decay rather than power-law behaviour, and concluded that even for a
semi-infinite metal inclusion of correlation is essential for the
correct
description of the image potential felt by unoccupied states. 

\subsection{Image-potential states}

An electron outside a metal surface can be trapped by the
image-potential
induced surface barrier, if its energy lies below the vacuum level and is
reflected from the metal due to the lack of allowed bulk states along
certain
symmetry directions. This situation occurs when for a given wave vector
parallel to the surface ${\bf k}_\parallel$ 
there is a gap in the projected bulk band structure.
Noble and transition metals usually present a surface band gap near the
vacuum
level $E_v$, and the long-range character of the image potential then
gives rise to a Rydberg-like series of bound image-potential induced
states whose binding energies converge for ${\bf k}_\parallel=0$ towards $E_v$
as dictated by Eq. (\ref{image}). In the absence of a band gap, these
image-states often persist as image resonances. 

More than two decades ago, Echenique and Pendry\cite{Echenique} carried
out a
theoretical investigation of the integrity of image states at metal
surfaces
by viewing these surface states as standing waves of an electron
bouncing back and forth between the semi-infinite solid and the surface
barrier. Image states at metal surfaces were first detected
experimentally by
LEED fine-structure analysis and then by inverse photoemission
(IP)\cite{Smith,Dose,Straub} and two-photon photoemission
(2PPE)\cite{Giesen,Fujimoto,Fauster2}. Although the Kohn-Sham
eigenfunctions and eigenvalues have no clear physical meaning, model xc
potentials that go beyond the LDA have been used to reproduce, within
DFT, the
experimentally observed binding energies and effective masses of image
states on a variety of metal
surfaces\cite{Weinert,Smith2,Nekovee1,Nekovee2}.

The finite lifetime of image states at metal surfaces, which is known to
be
mainly due to electronic excitations in the solid\cite{review}, is
inherently
outside the realm of DFT. In the framework of Green-function theory, one
identifies the inverse quasiparticle lifetime as follows
\begin{equation}\label{lifetime}
\tau_s^{-1}=-2\,{\rm Im}E_s,
\end{equation}
where $E_s$ is the quasiparticle energy. Assuming that the
eigenfunctions
$\phi_s(\br)$ and eigenvalues $\varepsilon_s$ of a suitable
hermitian single-particle Hamiltonian $\hat H_0$ approach the actual
quasiparticle wave functions and energies, $E_s$ can be approximately
determined
from Eq. (\ref{quasienergy}). Introduction of Eq. (\ref{quasienergy})
into Eq.
(\ref{lifetime}) then yields, on the energy-shell ($Z_s=1$),
\begin{equation}\label{lifetime2}
\tau_s^{-1}=-2\int d\br\int d\br'\,\phi_s^*(\br)\,{\rm
Im}\Sigma(\br,\br';\varepsilon_s)\,\phi_s(\br).
\end{equation}

If one further assumes translational invariance in the plane of the
surface, single-particle wave functions are of the form
\begin{equation}
\phi_s(\br)={1\over\sqrt A}\,\phi_s(z)\,{\rm e}^{i\,{\bf
k}_\parallel\cdot\br_\parallel},
\end{equation}
and the inverse quasiparticle lifetime is obtained as follows
\begin{equation}\label{lifetime3}
\tau_s^{-1}=-2\int dz\int dz'\,\phi_s^*(z)\,{\rm
Im}\Sigma(z,z';{\bf k}_\parallel,\varepsilon_s)\,\phi_s(z),
\end{equation}
where $\phi_s(z)$ and $\varepsilon_s$ now are image-state wave functions
and
energies describing motion normal to the surface, and $A$ represents the
normalisation area. 
  
The first quantitative evaluation of the lifetime of image states, as
obtained from Eq. (\ref{lifetime3}), was reported by Echenique {\it et
al.}\cite{Echenique2}. In this calculation, the image-state wave
functions $\phi_s(z)$ were approximated by hydrogenic-like wave functions with no
penetration into the solid and a simplified free-electron-gas (FEG)
model was used to approximate the electron self-energy. In subsequent
calculations the penetration of the image-state wave
function into the crystal was allowed\cite{Deandres}, thereby accounting for
this new decay channel. The role that the narrow crystal-induced Shockley
surface state on the (111) surfaces of Cu and Ni plays in the decay of the
$n=1$ image state on these surfaces was investigated by Gao and
Lundqvist\cite{Gao}. In this work, the image-state wave functions were also
approximated by hydrogenic-like wave functions with no penetration into the
solid, a simplified parameterised form was used for the Shockley surface-state
wave function, and screening effects were neglected altogether.

The first self-consistent many-body calculations of image-state lifetimes on
noble and simple metals were reported only very recently\cite{Chulkov1}, and
good agreement with experimentally determined decay times\cite{exp1,exp2,exp3}
was found. In Ref.\onlinecite{Chulkov1}, the decay rate of image states was
computed from Eq. (\ref{lifetime3}) with the electron self-energy evaluated in
the $G^0W^0$ approximation. The image-state wave function $\phi_s(z)$ and energy
$\varepsilon_s$ and all the eigenfunctions and eigenvalues entering the
non-interacting Green function $G^0(z,z';{\bf k}_\parallel,\varepsilon)$ were
obtained by solving a one-particle Schr\"odinger equation with a realistic
one-dimensional model potential, and the potential variation in the plane
parallel to the surface was considered thorugh the introduction of the
effective mass.

Self-consistent calculations of the key role that the Shockley surface
state plays in the decay of image states on Cu surfaces were later carried
out\cite{Chulkov2}, and the inclusion of short-range xc effects
was investigated\cite{Chulkov3} in the framework of the so-called $GW\Gamma$
approximation\cite{Rice,Mahan}. In this approximation, the electron self-energy
is of the $GW$ form, i.e., it is given by Eq. (\ref{gw}), but with an effective
screened interaction
\begin{equation}\label{wnew}
W({\bf r},{\bf r}';E)=v({\bf r}-{\bf r}')+\int{\rm d}{{\bf
r}_1}\int{\rm d}{{\bf r}_2}\,\left[v({\bf r}-{\bf
r}_1)+f_{xc}(\br_1,\br_2;E)\right]\,
\chi({\bf r}_1,{\bf r}_2,E)\,v({\bf r}_2-{\bf r}'),
\end{equation}
the density-response function now being the time-ordered counterpart of the
retarded density-response function of Eq. (\ref{chi}). The xc kernel
$f_{xc}(\br_1,\br_2;E)$ entering Eqs. (\ref{chi}) and (\ref{wnew}) account
for the reduction in the e-e interaction due to the existence of
short-range xc effects associated to the image-state electron and to screening
electrons, respectively. In Ref.\onlinecite{Chulkov3} the {\it exact} xc kernel
was replaced by that of Eq. (\ref{kernellda}), and it was concluded that
$G^0W^0$ calculations produce decay rates that are within $5\%$ of the more
realistic $G^0W\Gamma$ calculations. Although the presence of short-range
exchange and correlation between screening electrons significantly
enhances the decay probability of image states, this enhancement was found to
be more than compensated by the large reduction in the decay rate produced by
the presence of a xc hole around the image-state electron itself.

Unlike image states, which are unoccupied states with energies close to the
vacuum level, crystal-induced Shockley surface states are known to exist near
the Fermi level in the $\Gamma$-L projected bulk band gap of the (111) surfaces
of the noble metals Cu, Ag, and Au\cite{Gartland,Plummer,Himpsel}.
Hence, these surface states form a quasi two-dimensional (2D) electron
gas, which overlaps in energy and space with the three-dimensional (3D)
substrate, and represent a promising playground for lifetime investigations.

Photohole lifetimes of Shockley surface states were investigated in a variety of
metal surfaces with the use of high-resolution angle-resolved photoemission
(ARP) spectroscopy\cite{ARP1,ARP2,ARP3}. Recently, the STM was used to
determine the lifetime of excited holes at the edge of the partially occupied
surface-state band on the (111) face of noble metals\cite{stm1}, and also to
measure the lifetime of hot surface-state and surface-resonance electrons, as
a function of energy\cite{Burgi}. $G^0W^0$ calculations\cite{Science},
as obtained from Eq. (\ref{lifetime2}), demonstrated that the decay of
surface-state holes is dominated by 2D electron-electron interactions screened by
the underlying 3D electron system, and showed an excellent agreement with
the experiment. These theoretical investigations were then extended to the case
of hot surface-state and surface-resonance electrons in Cu(111)\cite{pss},
showing that, contrary to the case of surface-state holes, major
contributions to the e-e interaction of surface-state electrons above the Fermi
level come from the underlying bulk electrons and thereby giving an
interpretation to the measurements reported in Ref.\onlinecite{Burgi}.

\section{Perspectives for future research}
\label{sec:perspectives}

Approximate density-functional treatments of many-body effects
are being applied to increasingly more complex 
problems in surface science \cite{bird}. The question of their accuracy
in describing these effects, e.g., in chemical reactions at surfaces,
and how to improve them systematically is still outstanding. 
In the case of a jellium surface many-body effects 
are not masked by the effect of the crystal structure, so this system provides
an stringent test of the accuracy of approximate density functionals in
strongly inhomogeneous systems. In principle,  approximate density
functionals, as well as other many-body approaches, could be
benchmarked for the jellium surface against diffusion Monte Carlo results.
However, the presence of rather large statistical fluctuations 
(in the electron-density profile) and finite-size errors (in the surface
energy) makes it difficult to give  a clear verdict on the performance of LDA,
GGA, and RPA on the basis of the current DMC data. There
is  a clear need for {\it new} DMC calculations of the work function and the 
surface energy of jellium surfaces which are performed with the use of larger
system sizes, longer runs, and an improved treatment of finite-size effects. 

Although computationally much more expensive than DFT calculations, 
QMC algorithms have the important computational feature of being 
inherently parallel. The use of massively parallel computers
has made the application of QMC methods to the investigation of real solids
possible. Furthermore, QMC studies of real surfaces are now within reach, and a
full-many body description of chemical reactions at surfaces appears to be an
exciting  perspective further down the line.

Recently, the Green-function theory, within the GW approximation, has been
shown to provide total energies of an infinitely extended free-electron system
that are as accurate as those obtained from DMC calculations, as long as the
GW approximation is treated self-consistently\cite{Holm}. Within this approach,
the Green function entering the GW self-energy of Eq. (\ref{gw}) is evaluated
self-consistently from the Dyson equation [Eq. (\ref{Dyson})], the screened
interaction being obtained as in the RPA through Eqs. (\ref{w})-(\ref{g0}) but
with the non-interacting Green function $G^0(\br,\br';E)$ replaced by its
self-consistent interacting counterpart\cite{Holm,Eguiluzsc}. The total energy is
then obtained from the so-called Galitskii-Migdal formula\cite{Galitskii}. The
application of this way of obtaining the total energy to the case of a jellium
surface seems to be promising in the effort to understand the current
discrepancies between DFT and wave-function-based estimates of the surface
energy. 

Excited-state properties are still the realm of Green function 
methods (in the GW approximation). In the case of image-potential induced 
states these methods have the great advantage of providing, {\em within the same
computational framework}, a consistent description of both binding energies and
lifetimes, as well as the surface barrier. So far, full GW calculations have been
limited to the study of jellium surfaces and the (111) surface of Al, both of
which support only image resonances. However, the binding energies and
intrinsic lifetimes of these {\em resonances} are not well-defined and are
difficult to compare with the experiment. On the other hand, currently available
GW calculations of well-defined image {\em states} on noble metal surfaces
have been carried out with the use of a model one-dimensional potential, and
first-principles many-body descriptions that are based on a complete
three-dimensional treatment of the band structure of the solid would be
desirable. Since binding energies of image states are very sensitive to the
surface barrier, comparison between these calculations and the available
experimental data should also help to resolve remaining questions, such as the
precise many-body origin of the image potential.

\section*{Acknowledgements}
The authors would like to thank Matthew Foulkes for useful discussions in
connection with this research. J.M.P. acknowledges partial support by the
University of the Basque Country, the Basque Unibertsitate eta Ikerketa Saila,
and the Spanish Ministerio de Educaci\'on y Cultura.

\begin{table}[h]
\begin{tabular}{lccc}
Surface   & LDA$^a$  & EXP$^b$   \\ \hline
Nb(110)   & 4.66 & 4.87 &     \\
Mo(110)   & 4.94 & 4.95 &     \\
Pd(111)   & 5.53 & 5.6 &     \\
AG(111)   & 4.67 & 4.74 &      \\
Ag(100)   & 4.43 & 4.64 &      \\ 
Ag(110)   & 4.23 & 4.52 &      \\
\end{tabular}
\caption{LDA and experimental values of the work function for low-index
crystal
faces of $4d$ metals, taken from $^a\!\!$
Ref.\protect\cite{methfessel}\protect\- and $^b\!\!$
Ref.\protect\cite{workfun_exp2}\protect, respectively}
\label{table:workfun}

\end{table}
\begin{figure}\noindent
\label{figure:hole}
    \begin{tabular}{@{}ccccl@{}}
	\epsfxsize=3.20in\epsfbox{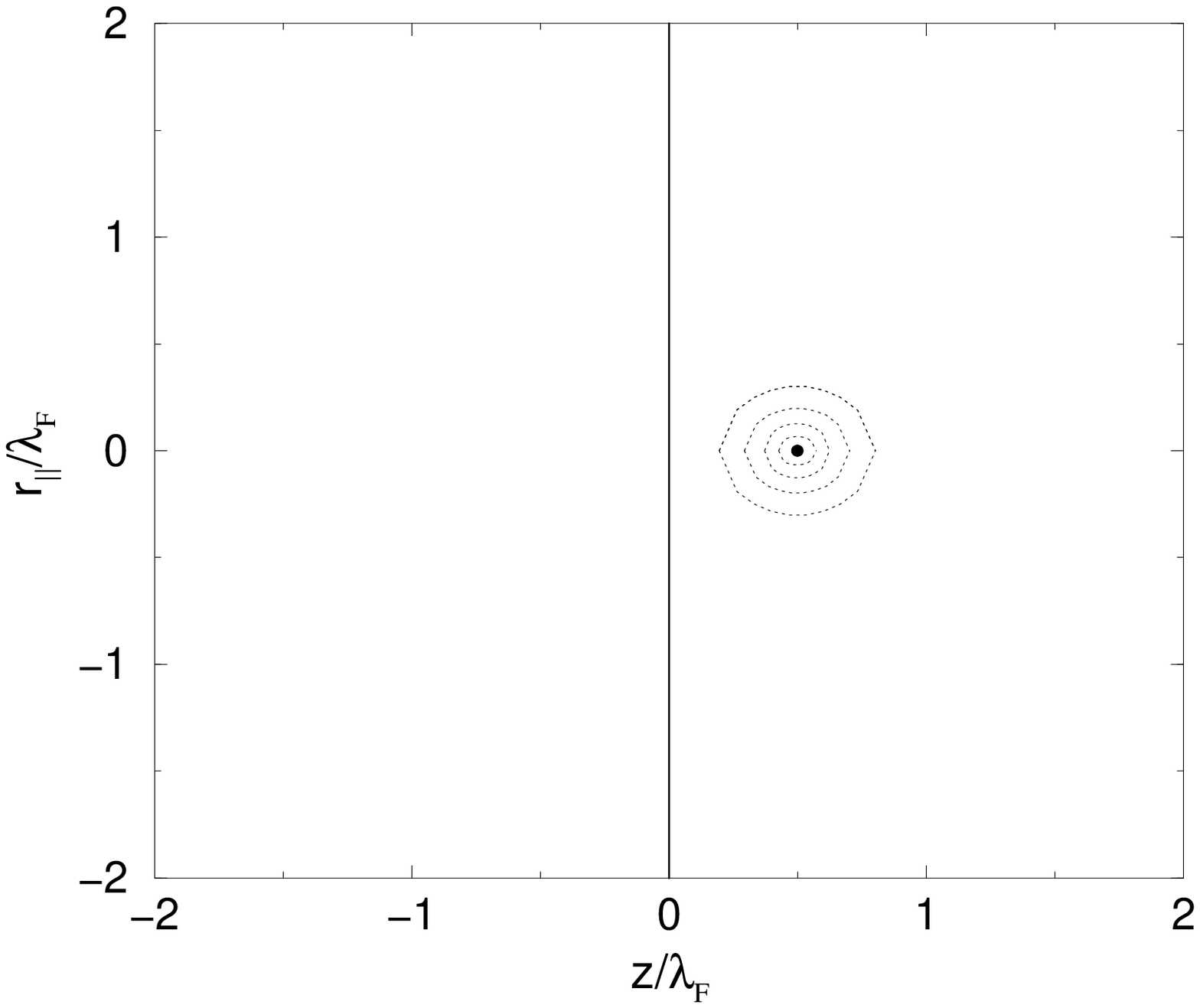}&
    	\epsfxsize=3.20in\epsfbox{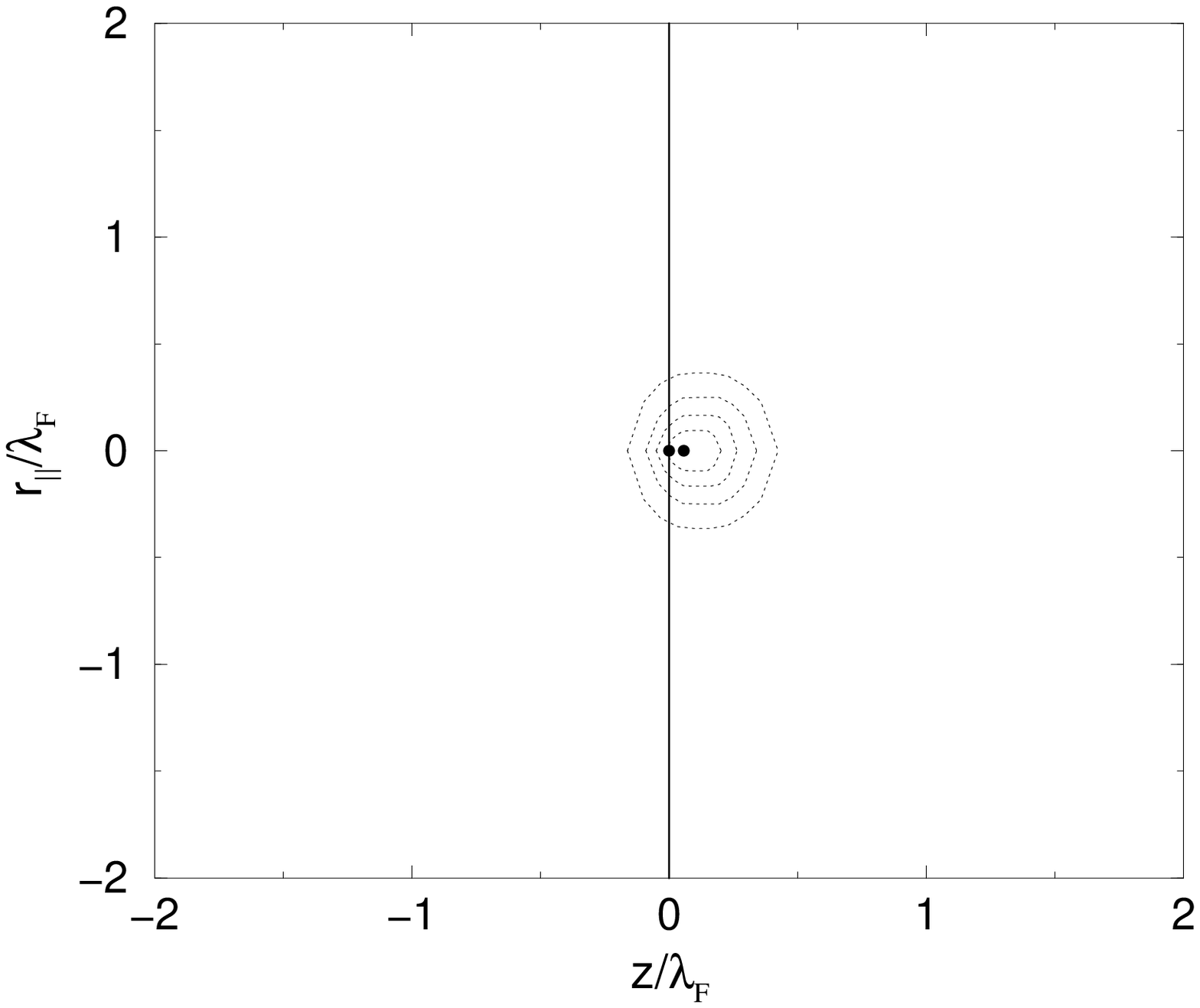}\\
    	\epsfxsize=3.20in\epsfbox{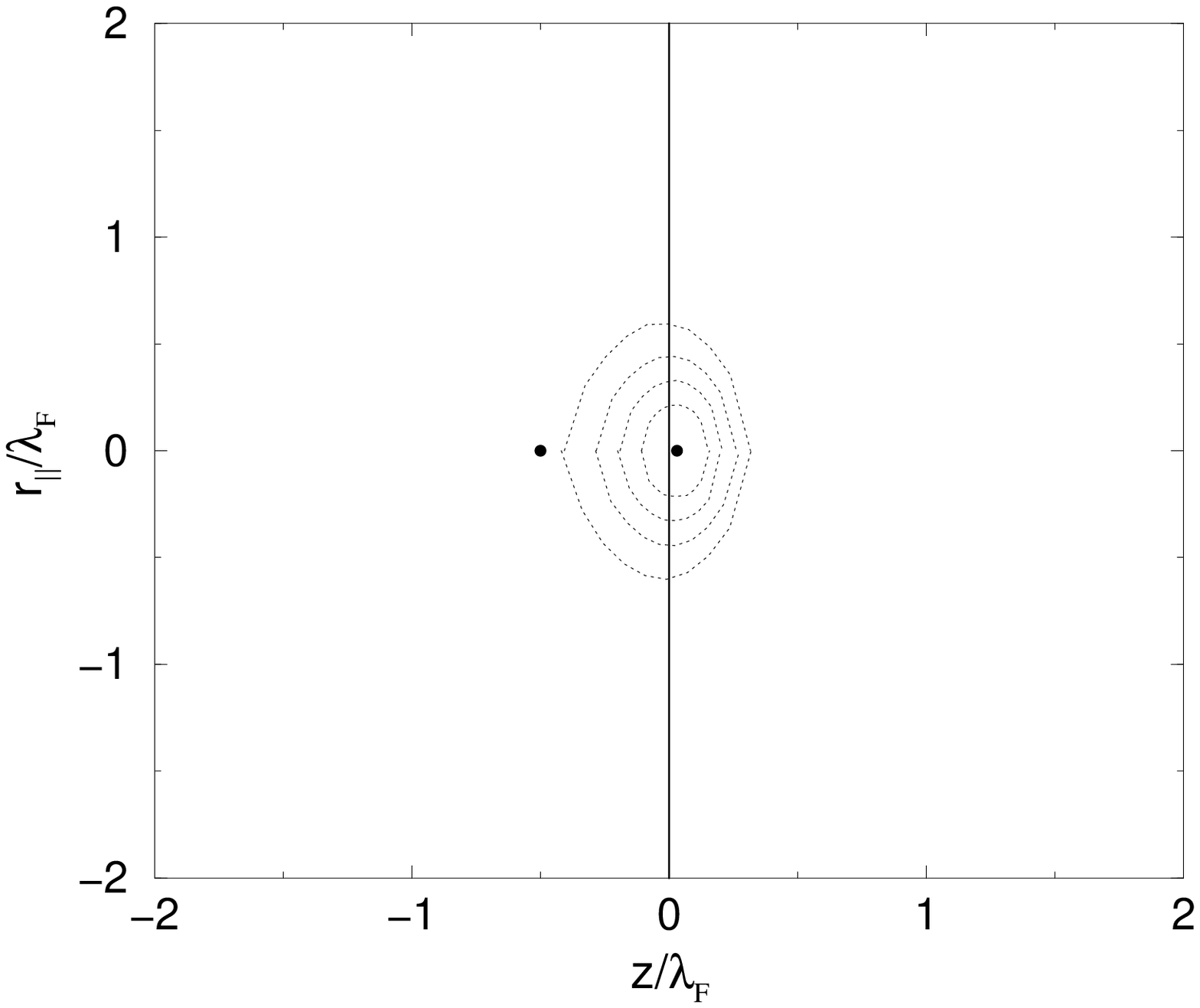}& 
    	\epsfxsize=3.20in\epsfbox{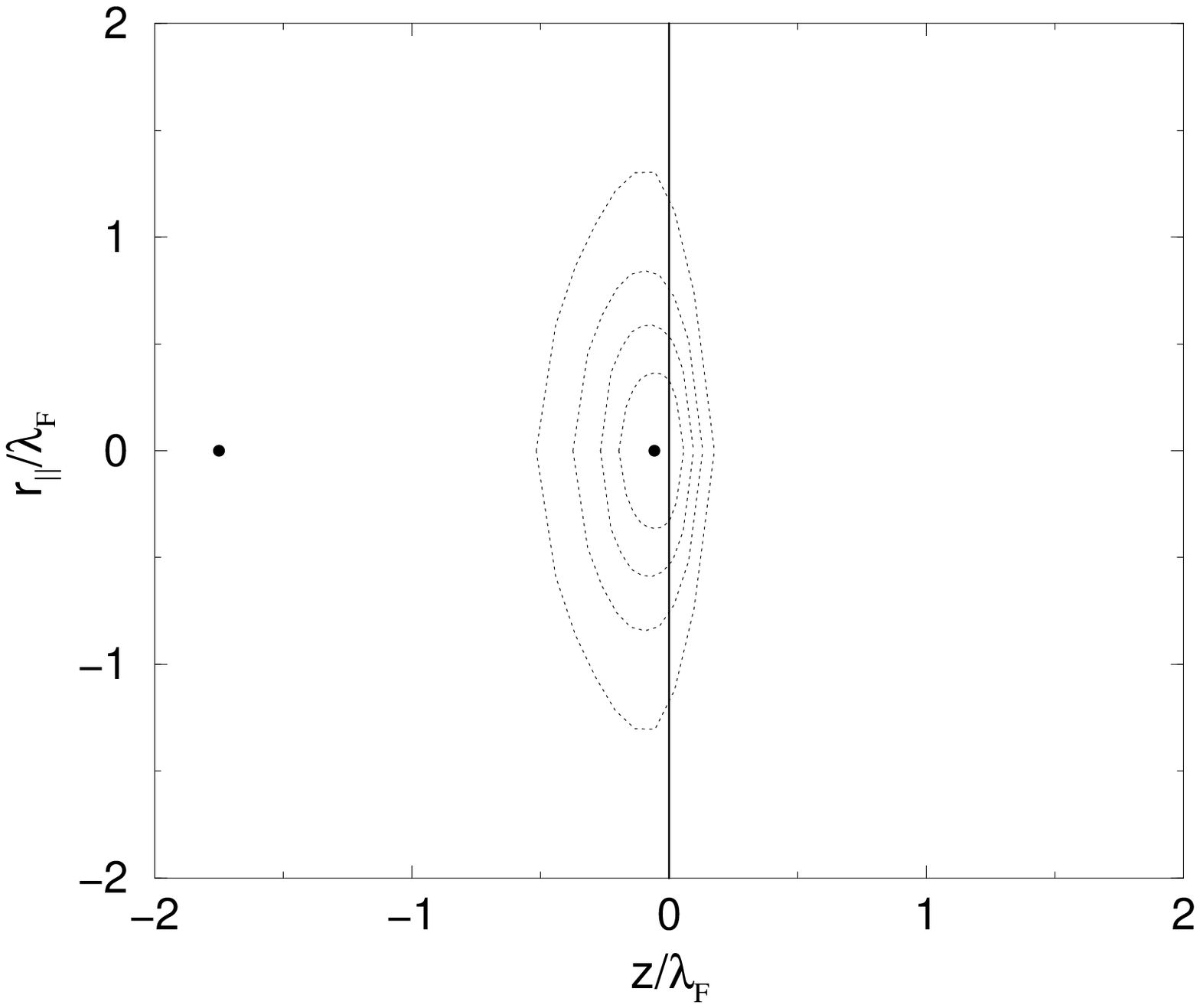}&
\end{tabular}
\caption{Contour plots of exchange-correlation hole for jellium 
surface for $r_s=2.07$ calculated within the DFT-based RPA.
Snapshots show deformation of the hole around  the electron 
as the electron moves from a point well inside metal to the vacuum.
The vertical line represents the edge of the positive background. One 
of the circles represent the position of the electron. The other circle 
represent the centre of gravity of the hole. Metal occupies $z\geq0$.}

\end{figure}

\end{document}